\documentclass[journal,onecolumn]{IEEEtran}
\usepackage{amsmath}
\usepackage{graphicx} 
\AtBeginDocument{         %
\providecommand\BibTeX{{  %
\normalfont B\kern-0.5em{\scshape i\kern-0.25em b}\kern-0.8em\TeX}}
}

\def\TP {\textrm{TP}}
\def\FN {\textrm{FN}}
\def\FP {\textrm{FP}}
\def\TN {\textrm{TN}}

\begin{document}
\title{Machine Learning-Based Automatic Cardiovascular Disease Diagnosis Using Two ECG Leads}
\author{Cheng Guo, Sajid Ahmed, {\em Senior Member, IEEE},  and Mohamed-Slim Alouini, {\em Fellow, IEEE}\\
King Abdullah University of Science and Technology (KAUST), Thuwal, KSA
gc466258802@berkeley.edu, sajid.ahmed@kaust.edu.sa, slim.alouini@kaust.edu.sa.}

\begin{abstract}
The state-of-the-art cardiovascular disease diagnosis techniques use machine-learning algorithms based on feature extraction and classification. In this work, in contrast to conventional single Electrocardiogram (ECG) lead, two leads are used, and auto-regressive (AR) coefficients and statistical parameters are extracted to be used as features. Four machine-learning classifiers support-vector-machine (SVM), $K$-nearest neighbors (KNN), multi-layer perceptron (MLP), and Naïve Bayes are applied on these features to test the accuracy of each classifier. For simulation, data is collected from the MIT-BIH and Shaoxing People’s Hospital China (SPHC) database. To test the generalization ability of our proposed methodology machine-learning model is built on the SPHC database and tested on the MIT-BIH database and self-collected datasets. In the single-database simulation, the MLP performs better than the other three classifiers. While in the cross-database simulation, the SVM-based model trained by the SPHC database shows superiority. For normal and LBBB heartbeats, the predicted recall respectively reaches 100\% and 98.4\%. Simulation results show that the performance of our proposed methodology is better than the state-of-the-art techniques for the same database. While for cross-database simulation, the results are promising too. Finally, in the demonstration of our realized system, all heartbeats collected from healthy people are classified as normal beats. 
\end{abstract}
\maketitle

\section{Introduction}
\begin{figure}[t]
\centerline{\includegraphics[width=2.5in]{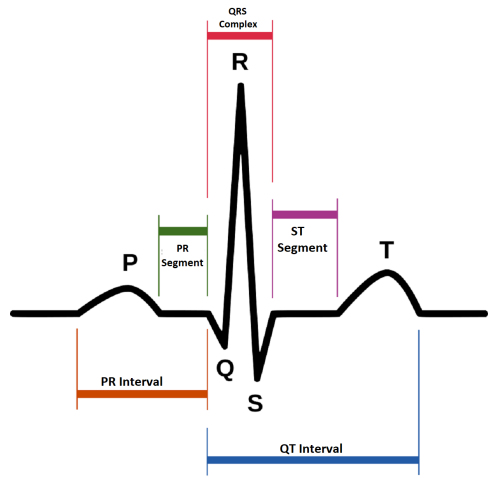}}
\caption{Morphology of a normal ECG signal. A normal ECG signal contains several peaks called P, Q, R, S, and T.}
\label{fig:ECG}
\end{figure}

The electrocardiogram (ECG) is an essential biological signal, which can reflect a person's heart conditions. Fig. \ref{fig:ECG} shows a simulated sample ECG signal of a normal person, which contains a QRS complex, two segments, and two intervals. Heart contraction continuously pumps blood throughout our body and creates the first heart electrical pulse on the sinoatrial node \cite{Tompkins_1993}. Then the electric pulse is conducted to the surface of our body that can be picked up by placing electrodes on the human's body to measure it.
The ECG signal's important role in diagnosing cardiac disorders is under research for over two decades. In traditional medical diagnosis, physicians analyze the ECG manually based on their experience. However, beat-by-beat arrhythmia detection and classification by a cardiologist can be very tedious and time-consuming. Therefore, automatic diagnosis of the heart condition is the need of the day, especially in situations where continuous monitoring is crucial. Unfortunately, the continuous real-time monitoring of heart activity of multiple patients using the conventional approaches is impractical even with multiple cardiologists due to human fatigue. Therefore, to release the pressure on cardiologists, an automatic processor-based diagnosis is the need of the day. 
State-of-the-art computer-based classification includes three main steps: heartbeat detection, feature extraction, and classification \cite{Roopa_2017}. Most of the R peak detection algorithms are based on Pan-Tompkins method \cite{Pan_1985}. Researchers have put forward many feature extraction methods to get the most representative parameters for the heartbeat \cite{Jambukia_2015}. There are many commonly used features such as discrete-wavelet-transform (DWT) coefficient \cite{Can_2012}, statistic parameters, intervals between two R-peaks, Hermite coefficients \cite{Lagerholm_2000,Lin_2014,Afkhami_2016}, and waveform shape features \cite{Chazal_2003,Ka_2011,Chazal_2006}. Principle component analysis (PCA) is also used to eliminate the dimension of features and reduce the computational cost. After extracting effective features from ECG signal, classifiers are trained and tested. In recent years, machine-learning (ML) based classifiers such as support-vector-machine (SVM) \cite{Celin_2018,Osowski_2004,Jiang_2006}, multi-layer-perceptron (MLP) \cite{Jiang_2007,Ince_2009}, and decision tree \cite{Rodriguez_2005} are becoming popular. Some of the algorithm will be briefly discussed in Sec. \ref{Sec:Classification}.%

Our proposed algorithm also uses ML algorithms and relies only on statistical parameters and auto-regressive (AR) model coefficients to reduce the number of features and computational complexity. 
\\
\\
{\bf Our Contribution:}
\begin{itemize}
    \item Most researchers use only one lead signal to classify cardiovascular disease (CVD) in the literature. In contrast, in our proposed work, ECG signals from two channels are exploited to significantly improve the classification accuracy. 
    \item Our proposed ECG signal classification method is fast and straightforward.
    \item Most of the research uses only the MIT-BIH database \cite{Moody_2001} and classify the ECG signals into two to six CVDs. In contrast, we classify the MIT-BIH database into eight CVDs. Moreover, the proposed algorithms are also tested on recently introduced Shaoxing People's Hospital China (SPHC) database. 
    \item We propose a novel cross-database simulation method in which the ML model is trained using one database and tested on the other databases. 
\end{itemize}
The remainder of the paper is organized as follows. In Sec. \ref{Sec:Meth}, the problem is formulated and the proposed method is discussed. Next, in Sec. \ref{Sec:Classification} classifiers are discussed, and simulation results are dicussed in Sec. \ref{Sec:Sim}. Finally, concluding remarks are drawn in Sec. \ref{Sec:Con}.

\section{Problem Formulation}\label{Sec:Meth}
The best classifier can be defined as the one that yields the best performance with a minimum number of features. Conventional classifiers use statistical parameters of ECG signals as features. In \cite{Arjunan_2016}, mean, variance, standard deviation, and skewness of an ECG signal are used to model a support-vector-machine (SVM) classifier to classify a patient as normal or abnormal. The accuracy of this classifier was $90\%$. In \cite{Celin_2018}, the same statistical parameters have been used to design SVM, Adaptive Boosting, Naïve Bayes, and artificial-neural-network (ANN) based classifiers. Due to the small number of parameters used, the classification accuracy is far less than required in the medical field. To further improve the accuracy, in \cite{Sarkaleh_2012} 24 DWT coefficients with Daubechies wavelet are extracted and used as features in MLP for classification. The accuracy of this work reached $96.5\%$. In \cite{Can_2012}, DWT coefficients, PCA, and R-R interval are used to extract features. The accuracy of this methodology was 99.3\% in the “class-oriented” evaluation and $86.4\%$ in the “subject-oriented” evaluation. The authors in \cite{Zhao_2005} have used both DWT and AR coefficients for feature extraction and applied SVM for classification. It has shown pretty high performance, which reached $99.68\%$. However, the total number of features used is 36, which is higher than many other schemes. The MIT-BIH database contains ECG signals from only 47 patients. The other drawback of the this scheme is that the accuracy significantly degrades to 37.1$\%$ when applied to the SPHC database, which contains signals from more than 10000 patients. 
In state-of-the-art algorithms, most of the features are extracted from computationally complex DWT coefficients. In the recent work \cite{Saira_SR_2021}, to reduce the computational complexity, simple features of an ECG signal PR and RT interval along with age and sex of patients are exploited. With these features, SVM applied to the MIT-BIH database yields an overall accuracy of $98.4\%$ and for the SPH database $83.87\%$. The cross-database results of this algorithm are not very promising and require further research. 
The overall accuracy of some of the state-of-the-art algorithms trained and tested on the MIT-BIH database, along with the number of CVD  they diagnose, is shown in Table \ref{Tabel:Summary}.    
\newcommand{\tabincell}[2]{\begin{tabular}{@{}#1@{}}#2\end{tabular}} 
\begin{table*}[hbtp]
\renewcommand\arraystretch{1.0}
\centering
{\caption{Accuracy of the existing methods trained and tested on MIT-BIH database.}\label{Tabel:Summary}}
\resizebox{\textwidth}{!}
{
\begin{tabular}{|c|c|c|c|c|c|}
\hline  
\bfseries  \tabincell{c}{Algorithm\\ {No. of CVD Diagnose}}& \tabincell{c}{\bfseries Number of \\ \bf Features}  & \bfseries Features & \bfseries Classifier & \tabincell{c}{\bfseries Performance\\ \bf Measures} & \bfseries \\\hline  
\tabincell{c}{Arjunan et al., \cite{Arjunan_2016} \\ {\bf Two Arrythmias}} & 4	 & Statistics Parameters & 	SVM & \tabincell{c}{Sensitivity   \\ Specificity}  & \tabincell{c}{	90\% \\ 90\%} \\\hline  

\tabincell{c}{Sarkaleh et al., \cite{Sarkaleh_2012} \\ {\bf Three Arrythmias}} &	24 &	DWT &	MLP &	Accuracy &	96.5\% \\\hline 
\tabincell{c}{Zhao et al., \cite{Zhao_2005}\\{\bf Six Arrythmias} }&	36 & \tabincell{c}{DWT\\
AR}	&	 SVM &	Accuracy &	99.68\% \\\hline  
\tabincell{c}{Can et al., \cite{Can_2012}  \\ {\bf Sixteen Arrythmias}}&42 &	\tabincell{c}{DWT \\
ICA \\ RR}	& SVM	& Accuracy &	\tabincell{c}{ “Class-oriented” 99.3\%\\ 
 “Subject-oriented” 86.4\% } \\\hline  
\tabincell{c}{Celin et al., \cite{Celin_2018} \\ {\bf Two Arrythmias}}& 4 &		Statistical Parameters & \tabincell{c}{SVM\\ Adaboost\\ ANN \\ Naïve Bayes} & \tabincell{c}{Accuracy 
} & \tabincell{c}{87.5\% \\93\% \\ 94\% \\ 99.7\% }  \\\hline %
\end{tabular}}
\end{table*}
In contrast to the above algorithms, our proposed algorithm uses ECG signals from two lead and use the Pan-Tompkins algorithm to detect R peak. Once an R peak is found, four AR coefficients and four statistical parameters mean, variance, standard deviation, and skewness of each heartbeat are found. Therefore, for each channel, an eight-dimensional feature vector is formed. Since two leads data is used, the feature vector becomes sixteen dimensional. 

\subsection{Proposed Methodology}
\begin{figure}[t]
  \centering
  {\includegraphics[width=4.0in]{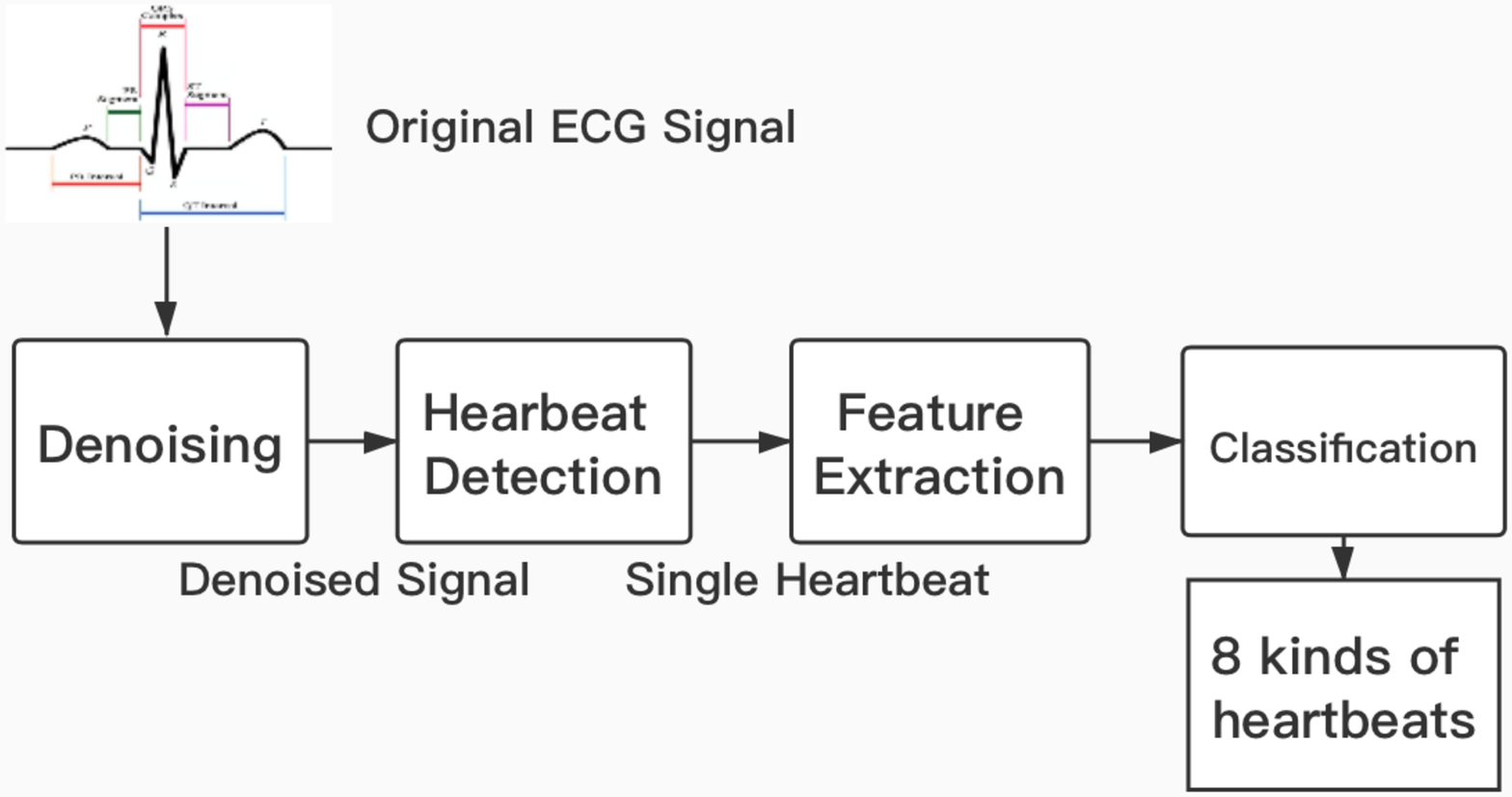}}
  {\caption{Proposed ECG classification block diagram. The raw ECG signal is denoised, features are extracted and passed to the machine learning algorithm for the classification of cardiovascular disease.}\label{fig:BlockDiagram}}
\end{figure}

\begin{figure}[h]
  \centering
  \includegraphics[width=3.5in]{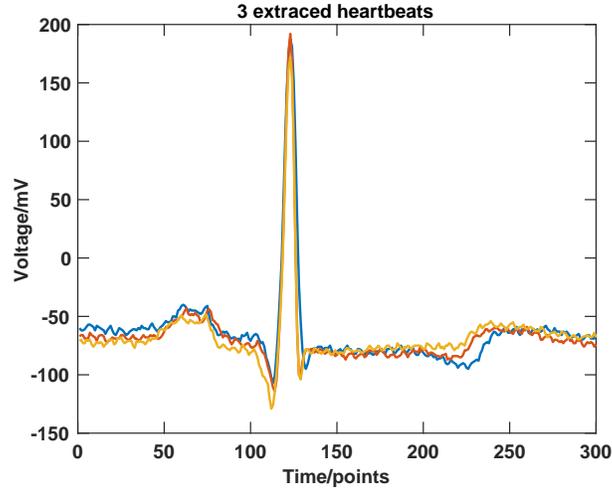}
  \caption{Three different heartbeat extracted from the database before denoising.}
  \label{fig:Tree_heartbeat}
\end{figure}

The block diagram of our proposed methodology is shown in Fig. \ref{fig:BlockDiagram}. It mainly consists of four stages: preprocessing, heartbeat detection, feature extraction, and classification. 
In the first stage, the original ECG signal is collected from the database, which is contaminated by the receiver noise and artifacts from other bio-electrical activities. Noise and artifacts can lead to low accuracy in the classification process. Therefore, they are removed to get as much as possible pure ECG signals. In our methodology, a simple median filter is used to remove the noise and artifacts. An example of three different raw heart beats are shown in Fig. \ref{fig:Tree_heartbeat}.
In the second step, P, QRS, and T waves of an ECG signal are detected. Classification can be done based on a single or whole ECG signal. In our methodology, R-peak of every heartbeat is found using the Pan-Tompkins algorithm proposed in \cite{Pan_1985}. After detecting the R peak, 122 points before and 177 points after the R peak are collected.  
In the third step, features from the ECG signal are extracted. Four AR coefficients,  along with four statistical parameters mean, variance, standard deviation, and Skewness of each ECG signal, are considered to reduce the feature vector dimension. These features can be easily calculated as
\begin{eqnarray}\label{eq:AR model}
x_n &=& \sum_{i=1}^{p}a_i x_{n-i} + e_n, \notag\\
\mu &=& \frac{x_1+x_2+\ldots + x_N}{N},\notag\\
\sigma &=& \sqrt{\frac{\sum_{i=1}^{n}\left(x_i - \mu\right)^2}{n}}, \notag\\
\mbox{Skewness} &=& \mbox{E}\left[\left(\frac{x_n -\mu}{\sigma}\right)^3\right],
\end{eqnarray}
where $x_n$ is the $n$th sample of ECG signal, $a_i$ is the $i$th AR coefficient, $p$ is the order of AR process, $e_n$ is the noise sample, $\mu$ is the mean value, and $\sigma$ is the standard deviation. In such a way, for each 300 points heartbeat, a feature vector of dimension eight is extracted. To improve the accuracy, we applied the feature extraction method to two different channels and concatenated them to form a 16-dimension feature vector. 

In the final step, these features are passed to the classifier. During this process, numerical properties of unknown signal features are analyzed, mainly in order to categorize them into different classes using training features of known classes. The classification procedure typically employs two phases: training and testing. The features of known samples and their classes are fed into the ML classifier during the training part. Then, the trained classifier is tested with the test sample. Based on the test outcomes with ground truth data, the performance of the system is analyzed.
\subsection{Datasets}
Two databases to test our algorithm are used. The first one is the MIT-BIH arrhythmia database, which is most extensively used in the literature to evaluate the performance of proposed algorithms. The MIT-BIH database contains 48 half-hour ECG signals from 47 subjects. It is recorded in two channels (leads), denoted as lead A and lead B. Signals are band-pass filtered at 0.1–100 Hz and sampled at 360 Hz. Out of 47, twenty-three records represent normal sinus rhythm (NSR), while the remaining 25 records include less common but clinically significant cardiac abnormalities. In 45 recordings, lead A is a modified-limb-lead II (MLL-II) signal and the lead B is usually a modified-limb-lead V1 (MLL-VI) (occasionally V2 or V5, and in one instance V4) signal. In the other three recordings, lead A is the V5 signal, and lead B is V2 (two instances) or MLII (one instance, i.e., signals were reversed). Each recording has an annotation file, which records the annotation detail, including the location of the QRS complex and the category of each heartbeat. Experts manually label it. In our single-database simulation part, the performance of the proposed approach is also tested on this database, which allows the direct comparison with already published results.

The other database used for the simulation is the SPHC database. This database contains 12 lead ECGs of 10646 patients with a 500Hz sampling rate. In addition, it includes 11 common rhythms and 67 additional cardiovascular conditions, all labeled by professional experts. The duration of each dataset is 10 seconds. A comparison of these two datasets is shown in Table \ref{tab:dataComp}:
\begin{table*}[t]
\centering
{
\caption{Number of subjects, signal duration of each subject, sampling rate, ages, and gender information in MIT-BIH and SPH databases.}\label{tab:dataComp}}
\renewcommand{\arraystretch}{1.0}
\resizebox{\textwidth}{!}
{
\begin{tabular}{|c|c|c|c|c|c|c|}
\hline  
\bfseries Database & \bfseries Subjects  & \bfseries Records (length) & \bfseries Sampling rate & \bfseries Age & \bfseries Male, n(\%) & \bfseries Number of Leads\\\hline  
MIT-BIH &	47 &	48 (30min) &360Hz	& 23-89 &	25 (52.08) &	2 \\\hline  
SPHC & 10646 & 10646 (10sec) &500Hz &4-98 &5956 (55.95) & 12   \\\hline   
\end{tabular}}
\end{table*}
\section{Classification} \label{Sec:Classification}
Four different ML classifiers SVM, KNN, MLP, and Naïve Bayes are used and briefly described in the following section to classify the features extracted from the ECG signal.
\subsection{Support Vector Machine Classifier}
\begin{figure}[htbp]
  \centering
  \includegraphics[width=3.5in]{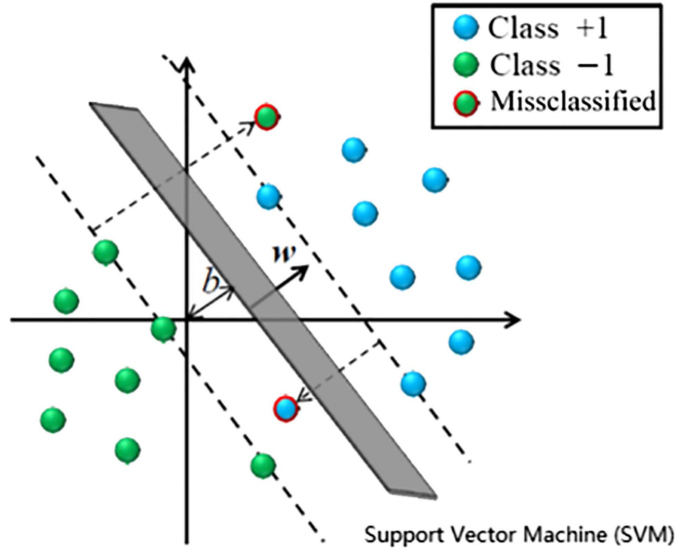}
  \caption{Basic principle of support vector machine, where input data is spread into two regions according to its class.}
  \label{fig:SVM}
\end{figure}

It is a class of supervised learning algorithms \cite{Vapnik_2013}. Traditional SVM is a generalized linear classifier that performs binary classification of data. The SVM algorithm maps training data points in the space in such a way that the distance between two categories is maximized as shown in Fig. \ref{fig:SVM}. New data is then mapped into the same space, and the category it belongs to is predicted based on some criterion. 
In our proposed methodology, SVM is a Multi-category SVM (MC-SVM). It classifies MIT-BIH dataset into 8 and SPHC dataset into 4 classes. We use the Gaussian radial basis function (RBF) as kernel to map original data into higher dimensions. Suppose $X=(x_1,...,x_d)$ is a set of features of dimension $d$, the SVM kernel can be defined as
\begin{equation}\label{eq:SVM}
K(x_i,x_j)=e^{-{\| x_i-x_j\|}^2/2\sigma^2}, 
\end{equation}
where $x_i$ and $x_j$ respectively represent the $i$th and $j$th feature of $X$.
The classification mainly consists of two processes: training and testing. After the R peak detection, each heartbeat segment is extracted, which contains 300 samples.

\subsection{K-Nearest Neighbor Classifier}
\begin{figure}[htbp]
\centering
\includegraphics[width=3.0in]{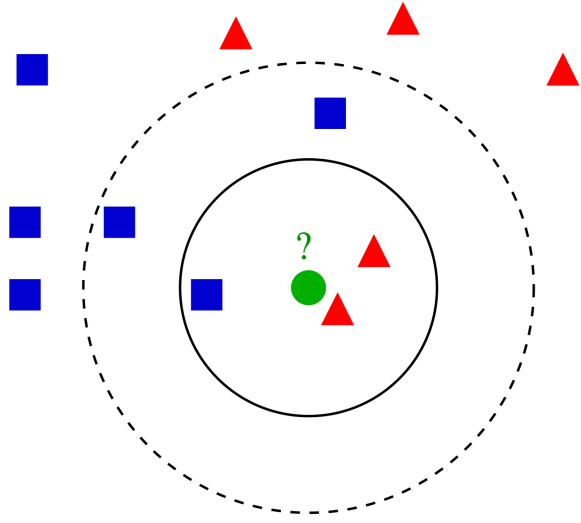}
\caption{Graphical representation of the KNN algorithm.}
\label{fig:KNN}
\end{figure}

The K-Nearest Neighbor (KNN) is a fundamental algorithm in ML \cite{Cover_1967}. The main idea behind KNN is that if $k$ nearest neighbor of a target are chosen, the class having more neighbors in the vicinity of the target will be the class of the target. The basic idea of the algorithm is presented in Fig. \ref{fig:KNN}. For example, if test sample is a green dot and $k=3$. Three nearest samples can be identified by drawing a solid circle, where two red-triangles and one blue-square can be seen. Therefore, according to the KNN classifier, the test sample will be classified as red-triangle. Similarly, if $k=5$, nearest neighbors can be identified by drawing a dash-line circle. Since there are three neighbors belonging to blue-squares and only two neighbors belong to red-triangle, the test point will be classified as blue-square. The distance between two samples $x_i$ and $x_j$ can be found using the following norm as 
\begin{equation}\label{eq:KNN}
L_p(x_i,x_j) = {\left(\sum_{l=1}^{d}{\left|x_i^{(l)}-x_j^{(l)}\right|}^p\right)}^\frac{1}{p}. 
\end{equation}
In (\ref{eq:KNN}), when $p=1$, the norm will be the Manhattan distance and for $p=2$ it will be the Euclidean distance. 

In this algorithm, the selection of $k$ is very important. Selecting small value of $k$ means less neighboring training instance to make a prediction. The approximation error of learning will be reduced, and only the training instance close to the input instance will play a role in the prediction result. However, the disadvantage is that the estimation error of learning will increase and the prediction results will be sensitive to the nearest instance points. 
On the other hand, selecting large value of $k$ will bring more training instances in the neighborhood for prediction. Its advantage is that the estimation error of learning can be reduced, but the approximate error will increase, that is, the prediction of the input instance is not accurate. 

\subsection{Multi-Layer Perceptron (MLP)}
\begin{figure}[htbp]
  \centering
  \includegraphics[width=3.25in]{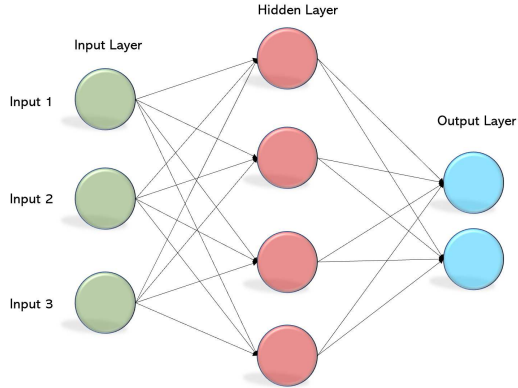}
  \caption{Basic model of multi-layer perceptron.}
  \label{fig:MLP}
\end{figure}

It is also called an artificial-neural-network \cite{Linh_2003}\cite{Hu_1993}. It is a feed-forward neural-network model. Besides input and output layers, it can have multiple hidden layers in the middle. The basic block diagram of the MLP is shown in Fig. \ref{fig:MLP}, where it can be seen that layers are fully connected to each other.
The simplest MLP has an input layer at the start, a hidden layer in the middle, and an output layer at the end. In particular, given a sample set  $X\in{\bf R}^{n\times d}$, where $d$ is the total number of features and $n$ is the total number of samples. Suppose MLP has only one hidden layer, the number of neurons is $h$, and the output of the hidden layer is $H\in{\bf R}^{n\times h}$. Since the hidden and output layer is fully connected, we denote weight and bias parameters of the hidden and output layers respectively by $W_h\in{\bf R}^{d\times h}$, $b_h\in{\bf R}^{1\times h}$, $W_o\in{\bf R}^{h\times q}$, and $b_o\in{\bf R}^{1\times q}$. Also denote activation functions for hidden and output layers by $g_h, g_o$. Then the output $O\in{\bf R}^{n\times q}$ can be calculated as
\begin{equation}\label{eq:MLP}
O = g_o(g_h(X W_h + B_h)W_o + B_o), \notag
\end{equation}
where
\begin{eqnarray}
B_{h} = \left(\begin{array}{c}b_{h} \\ b_{h} \\ \cdot \\ \cdot \\ \cdot \\ b_{h}\end{array}\right) \in {\bf R}^{n \times h},~ 
B_{o} = \left(\begin{array}{c}b_{o} \\ b_{o} \\ \cdot \\ \cdot \\ \cdot \\ b_{o}\end{array}\right) \in {\bf R}^{n \times q}.\notag
\end{eqnarray}
%
\subsection{Naïve Bayes Classifier}
This model is based on Bayes' principle and uses the knowledge of probability statistics to classify sample data set \cite{Padmavathia_2015}. It assumes that the attributes of a given target are conditionally independent of each other. This assumption greatly simplifies the complexity of Bayesian methods for practical scenarios. Consider a sample dataset $D = \{d_1,d_2,\ldots,d_n\}$ and corresponding feature set  $X=\{x_1,x_2,\ldots,x_d\}$. Class variables $Y={y_1,y_2,...,y_m}$. $x_1,x_2,...,x_d$ are independent and random. Then we can get the probability that a sample data belongs to class $y_i$ is:
\begin{equation}\label{eq:Naïve Bayes}
P(y_i|x_1,x_2,...,x_d) = \frac{P(y_i)\prod_{j=1}^{d}{P(x_j|y_i)}}{\prod_{j=1}^{d}{P(x_j)}}.
\end{equation}

We have used the ECG data from MIT-BIH Arrhythmia database corresponding to the normal heartbeat, and 7 types of arrhythmias: left bundle-branch-block (LBBB), right bundle-branch-block (RBBB), paced (PACE), premature-ventricular-contraction (PVC), atrial premature contraction (APC), fusion of ventricular and normal beat (FVNB) and fusion of paced-and-normal beat (FPNB). 
\section{Simulation Result} \label{Sec:Sim}
For simulation, to find the best value of SVM parameters $C$ and $\gamma$, grid search method is applied that yield $C=65536$ and $\gamma=2.44\times{10}^{-4}$. Similarly, for the MLP, after trying a different number of hidden neurons, a $16\times 100\times 50 \times 8$ network is used with two hidden layers. To compare the performance of our proposed classifier with the existing ones following performance metrices are used
\begin{eqnarray}
   \mbox{Overall~Accuracy} &=& \frac{\TP+\TN}{\TP+\TN+\FP+\FN} ,\notag\\
    \mbox{Precision} &=& \frac{\TP}{\TP+\FP}, \notag\\
    \mbox{Recall} &=& \frac{\TP}{\TP+\FN},\notag\\
f_{1}\mbox{-Score} &=& 2\frac{\mbox{Precision $\times$ Recall}} {\mbox{Precision $+$ Recall}} ,\notag
%
\end{eqnarray}
where TP denotes true-positive, FN denotes false-negative defined as the annotated peaks not detected by the algorithm, TN denotes true-negative defined as the patient had a CVD and the classifier also predicted patient is not normal, and FP denotes false-positive defined as peaks detected by the algorithm but not present. 

\begin{figure}[htbp]
  \centering
  \includegraphics[width=2.50in]{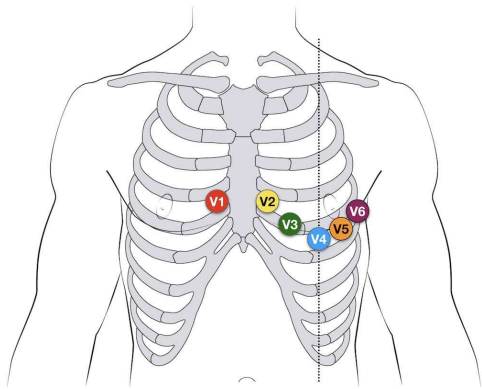}
  \caption{The location of different leads placement.}
  \label{fig:lead_place}
\end{figure}

To assess the performance of the proposed algorithms, real data from MIT-BIH and SPHC databases is used. In the MIT-BIH database, as mentioned earlier, MLL-II and MLL-VI channel signals are available for every dataset. The placement of V1 to V6 leads is shown in Fig. \ref{fig:lead_place}.
For eight types of heart conditions, normal, LBBB, RBBB, PACE, PVC, APC, FVNB, and FPNB, signals are extracted respectively from the records 100, 109, 118, 107, 208, 232, 213, 104 (combined with 217) of MIT-BIH database. From these records, 13000 labeled datasets are collected and split randomly into 70\% training and 30\% testing dataset. The above-discussed four learning models are trained and applied to test datasets for classification. The detailed results are shown in Table \ref{Table:MITBIH_SVM}, and compared with Zhao and Zhang's 32 feature algorithm \cite{Zhao_2005}. It can be seen in the table that the precision of all four ML models with only 16 proposed features is almost the same as Zhao and Zhang's algorithm which requires 32 features. 
%
\begin{table*}[htbp]
\renewcommand{\arraystretch}{1.5}
\centering
\caption{Precision comparison of Naïve Bayes, KNN, SVM, and MLP classifiers trained and tested on MIT-BIH database.}
\label{Table:MITBIH_SVM}
\renewcommand{\arraystretch}{1.5}
\resizebox{\textwidth}{!}
{
\begin{tabular}{|c|c|c|c|c|c|c|c|c|c|c|c|c|c|}
\hline  
          & \multicolumn{3}{c|}{\textbf{Naïve Bayes}} & \multicolumn{3}{c|}{\textbf{KNN}} & \multicolumn{3}{c|}{\textbf{SVM}} & \multicolumn{3}{c|}{\textbf{MLP}} &  \\
    \hline  
          & \textbf{Recall } & \textbf{F1-score} & \textbf{Precision} & \textbf{Recall } & \textbf{F1-score} & \textbf{Precision} & \textbf{Recall } & \textbf{F1-score} & \textbf{Precision} & \textbf{Recall } & \textbf{F1-score} & \textbf{Precision} & 
          \textbf{Precision in \cite{Zhao_2005}}
          \\
    \hline  
    \textbf{Normal} & 0.995 & 0.998 & 1.000 & 0.995 & 0.998 & 1.000 & 0.995 & 0.998 & 1.000 & 0.995 & 0.998 & 1.000 & 1.000 \\
    \hline  
    \textbf{LBBB} & 0.976 & 0.987 & 0.999 & 0.989 & 0.994 & 0.999 & 0.993 & 0.995 & 0.997 & 0.995 & 0.996 & 0.997 & 0.997 \\
    \hline  
    \textbf{RBBB} & 0.997 & 0.993 & 0.989 & 0.994 & 0.995 & 0.997 & 0.997 & 0.998 & 0.998 & 0.997 & 0.998 & 0.998 & 1.000 \\
    \hline  
    \textbf{PACE} & 0.968 & 0.975 & 0.981 & 0.994 & 0.981 & 0.969 & 0.987 & 0.987 & 0.987 & 0.987 & 0.984 & 0.981 & 0.997 \\
    \hline  
    \textbf{PVC} & 1.000 & 0.991 & 0.983 & 1.000 & 1.000 & 1.000 & 1.000 & 1.000 & 1.000 & 1.000 & 1.000 & 1.000 & 1.000 \\
    \hline  
    \textbf{APC} & 0.998 & 0.997 & 0.995 & 0.993 & 0.989 & 0.984 & 1.000 & 0.994 & 0.989 & 1.000 & 0.997 & 0.993 & 0.989 \\
    \hline  
    \textbf{FVNB} & 0.999 & 0.988 & 0.978 & 0.991 & 0.991 & 0.991 & 1.000 & 0.999 & 0.997 & 0.997 & 0.996 & 0.996 & 1.000 \\
    \hline  
    \textbf{FPNB} & 1.000 & 1.000 & 1.000 & 1.000 & 1.000 & 1.000 & 1.000 & 1.000 & 1.000 & 1.000 & 1.000 & 1.000 & 1.000 \\
    \hline  
    \end{tabular}
    }%
  \label{tab:addlabel}%
\end{table*}%
The overall accuracy of all four classifiers is compared with Zhao and Zhang's algorithm and shown in Table \ref{Tabel:AllAccuracy_MIT-BIH}. Table shows that although our proposed methodology uses only 16 features compared to the 36 features Zhao and Zhang's algorithm, the overall accuracy of both algorithms is very much close to each other. The overall accuracy of all classifiers using a single channel (lead VII) with only eight features is also shown in the last column of Table \ref{Tabel:AllAccuracy_MIT-BIH}. It can be seen here that although with a single lead, the overall accuracy of the algorithms with proposed features is slightly less but is not very far from Zhao and Zhang's algorithm. 
\begin{table}
\centering
  {\caption{Overall accuracy comparison of MLP, SVM, KNN, and Naïve Bayes models trained and tested on MIT-BIH database.} \label{Tabel:AllAccuracy_MIT-BIH}}
  {\begin{tabular}{|c|c|c|c|}
  \hline  
  \bfseries Classifier Model & \bfseries Proposed Method & \bfseries Zhao $\&$ Zhang & \bfseries Single Lead \\
                       &                           &    \cite{Zhao_2005}        &   (MLII)\\
  \hline   
  MLP & 99.7\% & 99.7\% & 99.4\% \\
  \hline 
  SVM & 99.6\% & 99.8\% & 99.2\% \\
  \hline 
  KNN & 99.3\% & 99.6\% & 98.7\% \\
  \hline 
  Naïve Bayes & 99.1\% & 99.1\% & 98.3\% \\
  \hline  
  \end{tabular}}
\end{table}
%
The proposed methodology is also tested on the SPHC database \cite{Zheng_2020} to see its generalization. The SPHC database contains 11 kinds of arrhythmia signals. However, some arrhythmia signals that include only tens of datasets are not included in the CVD classification. Therefore, we chose four main arrhythmia signals sinus bradycardia (SB), sinus rhythm (SR), atrial fibrillation (AFIB), and sinus tachycardia (ST). For feature extraction, V4 and V5 lead signals are used. The total number of heartbeats is 109109. Out of which 70\% are used for training and 30\% for testing data. For comparison, Zhao and Zhang's method is also applied to these datasets. Results of our proposed methodology and Zhao and Zhang's method are compared in Table \ref{Table:CompSPH}. It can be seen in the table that our methodology with only sixteen features outperforms the Zhao and Zhang's algorithm, which uses 36 features. 
The precision of the proposed methodology is also assessed on individual CVDs in the SPHC database and compared with the Zhao and Zhang's algorithm. However, to save space and time, only an MLP classifier is used. The corresponding performances are given in Table \ref{SPH_MLP}, where it can be seen that for almost all individual CVDs, the performance of the proposed methodology outperforms the Zhao and Zhang's algorithm. 
\begin{table}[hbtp]
\centering
{\caption{Overall accuracy comparison of MLP, SVM, KNN, and Naïve Bayes models trained and tested on SPH database.}\label{Table:CompSPH}}
{\begin{tabular}{|c|c|c|}
\hline  
\bfseries Classifier Model & \bfseries Proposed Algorithm & \bfseries Zhao $\&$ Zhang \cite{Zhao_2005}  \\ \hline  
   MLP         &  91.3\%  &  90.5\% \\ \hline  
   SVM         &  89.7\%  &  91.2\% \\ \hline  
   KNN         &  89.3\%  &  88.6\% \\ \hline  
   Naïve Bayes &  87.1\%  &  86.7\% \\
  \hline 
  \end{tabular}}
\end{table}
\begin{table}[hbtp] 
\centering
{\caption{Precision comparison of MLP model trained and tested on SPHC database.}\label{SPH_MLP}}
{\begin{tabular}{|c|c|c|c|c|}
\hline  
\bfseries Disease & \bfseries Recall & \bfseries F1-score & \bfseries Proposed Algorithm Precision & \bfseries Zhao $\&$ Zhang Precision \\ \hline SB    &	 0.912  & 	0.938  &  0.966	 &  0.897 \\ \hline  
SR	  &	 0.913  &   0.904  &  0.895  &  0.912 \\ \hline
AFIB  &	 0.896  &	0.900  &  0.905  &  0.892 \\ \hline 
ST    &	 0.952  &	0.933  &  0.915  &	0.895 \\ \hline  
\end{tabular}}
\end{table}

Generally, algorithms in the literature work fine if the testing datasets come from the same database used to train the model, as shown in the above simulations for MIT-BIH and SPHC databases. However, if the testing datasets are from an ECG machine with different specifications from the one used to build the original database, the ML algorithm may not yield satisfactory results. Building a database from scratch for each new ECG machine is not practical as it requires a large number of patients for each CVD and can be too time-consuming. Therefore, there is a need to develop a generalized ML model with specific features that can work with different ECG machines with slight tweaking.    

\begin{figure}[htbp]
  \centering
  \includegraphics[width=3.5in]{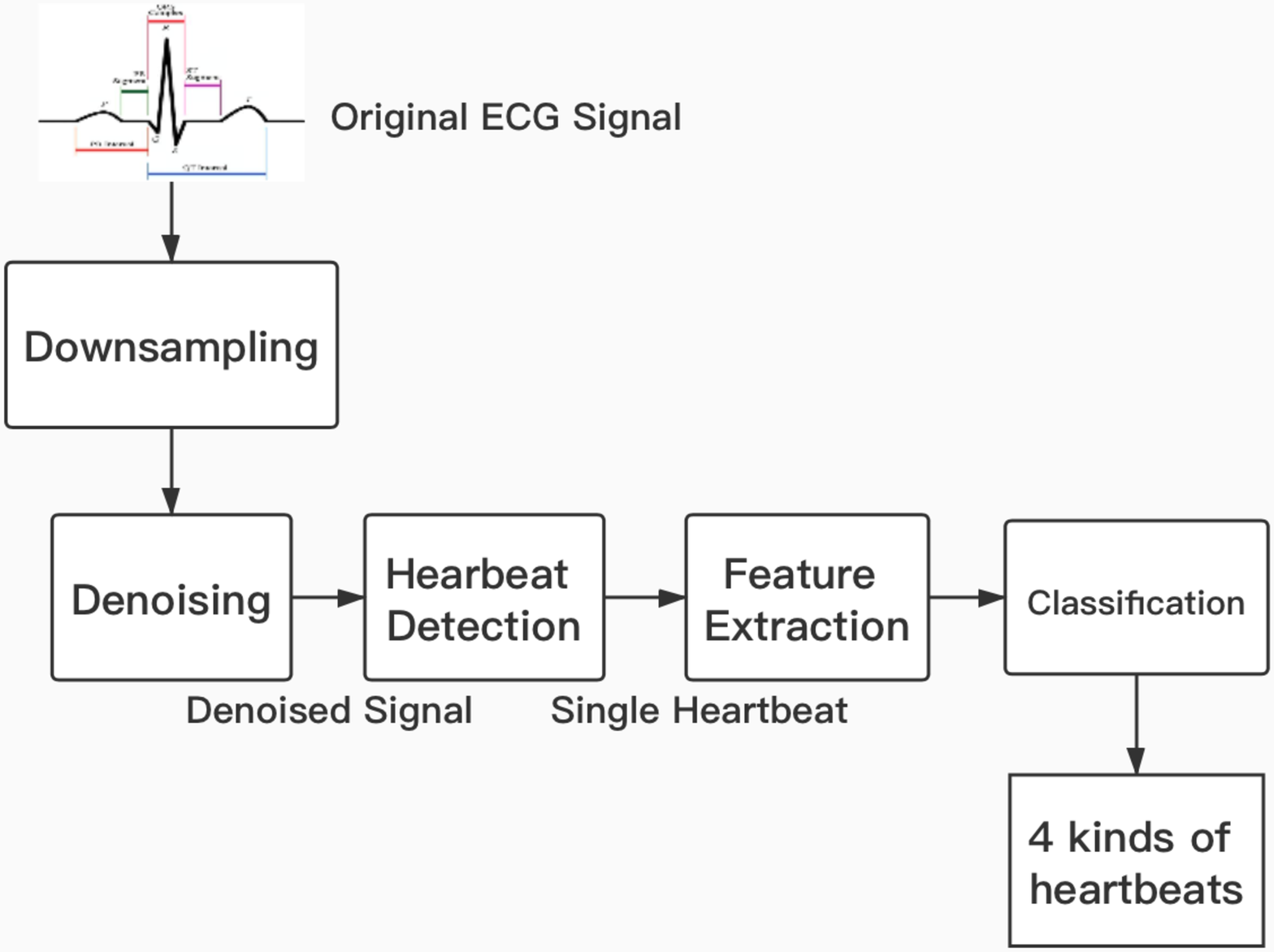}
  \caption{System design for cross-database classification.} 
  \label{Fig:CrossModel}  
\end{figure}

The main purpose of the following work is to test the proposed ML model on cross databases \cite{Georgoulas_2009}. Therefore, we train an ML model on one database and test it on other databases. For this, both databases should have the same sampling frequencies. If sampling frequencies are not the same, the testing dataset should be resampled at the sampling frequency of the training dataset \cite{Khawaja_2012}. In such situations, the selection of features and ML model becomes very important.
Therefore, we choose four CVDs signals normal, LBBB, RBBB, and APC, for cross-database simulation. They are recorded both in the MIT-BIH and SPHC databases. In the first step, we train an ML model using the MIT-BIH database and test it on datasets in the SPHC database after down-sampling to 360 samples per second. From what we observed, the built model could not accurately classify any heartbeats in the SPHC database. This is because there are thousands of heartbeats in the training dataset, but most come only from four patients. In addition, human ECG signals differ a lot from person to person. Therefore, it can be expected that the built model can’t predict patients in the SPHC database. Therefore, the ML model is trained with the SPHC database after down-sampling it to 360 samples per second in the second step. Finally, the built model is applied to the datasets of the MIT-BIH database for testing. The overall system design for cross-database simulation is shown in Fig. \ref{Fig:CrossModel}. 

We extract Normal, APB, RBBB, and LBBB signals present in MIT-BIH and SPM databases for cross-database simulations. For these CVDs, to train the model, 77137 heartbeats are extracted from V2 and V5 leads of the SPHC database to extract features. Out of which 70\% are used to train and 30\% to test the model. The Recall, F1-score, and precision using Naïve Bayes, KNN, and SVM classifiers with eights features extracted from each lead are shown in Table \ref{Table:CrossModel_SVM}. The simulation results show that the SVM classifier yields the best performance. Therefore, SVM is used to predict datasets in MIT-BIH database, which are shown in Table \ref{Table:Actual-Predicted}. It can be seen in the table that the proposed classifier's performance is satisfactory for Normal, LBBB, and RBBB but poor for APB. The SVM classifier can detect normal ECG signals with 100\% accuracy, LBBB with more than 98\% accuracy, RBBB with 86\% accuracy, and APB  with 32\% accuracy. The low accuracy of APB signals is due to the limited datasets.     

\begin{table*}[htbp]
\centering
\caption{Precision of Naïve Bayes, KNN, and SVM classifiers trained and tested on SPHC database}
\label{Table:CrossModel_SVM}
\renewcommand{\arraystretch}{1.0}
\resizebox{\textwidth}{!}
{
\begin{tabular}{|c|c|c|c|c|c|c|c|c|c|c|}
\hline
& \multicolumn{3}{c|}{\textbf{Naïve Bayes}} & \multicolumn{3}{c|}{\textbf{KNN}} & \multicolumn{3}{c|}{\textbf{SVM}} &  \\
\hline
& \textbf{Recall } & \textbf{F1-score} & \textbf{Precision}   & \textbf{Recall } & \textbf{F1-score} & \textbf{Precision}  & \textbf{Recall } & \textbf{F1-score} & \textbf{Precision} & \textbf{Support} \\
\hline
\textbf{Normal} & 0.958 & 0.908 & 0.863 & 0.981 & 0.921 & 0.868 & 0.995 & 0.933 & 0.878 & 19667 \\
\hline
\textbf{APB} & 0.017 & 0.032 & 0.254  & 0.041 & 0.070 & 0.224 & 0.018 & 0.035 & 0.576 & 1067 \\
\hline
\textbf{LBBB} & 0.272 & 0.227 & 0.195  & 0.162 & 0.249 & 0.535 &  0.678 & 0.696 & 0.716 & 475 \\
\hline
\textbf{RBBB} & 0.042 & 0.066 & 0.144  & 0.130 & 0.199 & 0.430 & 0.142 & 0.238 &0.733 & 1933 \\
\hline
\end{tabular}
}%
\end{table*}%
\begin{table}[h]
\centering
{\caption{Actual and predicted number of CVDs with SVM classifier trained on SPHC and tested on MIT-BIH database.} \label{Table:Actual-Predicted}}
{\begin{tabular}{|c|c|c|}
\hline  
\bfseries CVD & \bfseries Actual Number   & \bfseries Predict Number  \\ \hline  
Normal &  2237  & 2237    \\ \hline   
LBBB   &  2490  & 2450 (APB=1, Normal = 39)     \\ \hline   
RBBB   &  2164  & 1859 (Normal = 305)     \\ \hline 
APB    &  734   & 23 (Normal = 425, RBBB = 286)     \\ \hline 
  \end{tabular}}
\end{table}
\subsection{Self-Collected Data} \label{sec.ps}
\begin{figure}[htbp]
\begin{minipage}[t]{0.20\linewidth}
\centering
\includegraphics[height=4.5cm]{9.eps}
\end{minipage}%
\begin{minipage}[t]{0.5\linewidth}
\centering
\includegraphics[height=4.5cm]{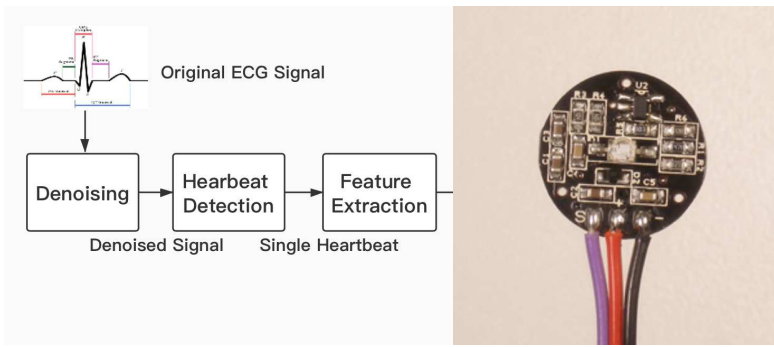}
\end{minipage}
\caption{Front and backside of the pulse sensor.} \label{fig:FrontBackSensor}
\end{figure}

\begin{figure}[htbp]
  \centering
  {\includegraphics[width=3.6in]{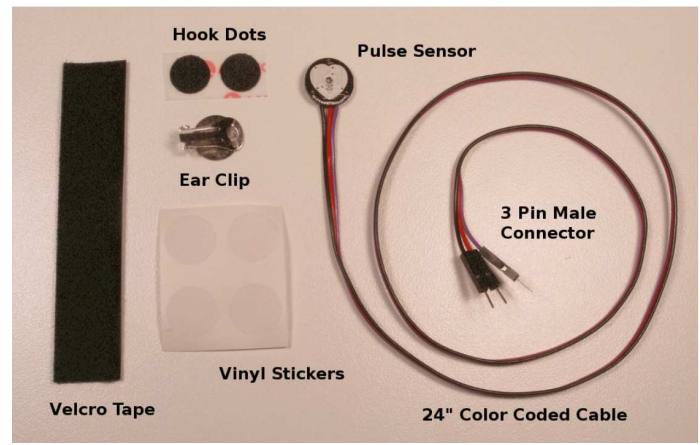}}
  {\caption{Pulse sensor kit.}\label{fig:FrontBackSensor2}   }
\end{figure}

\begin{figure}[htbp]
  \centering
  {\includegraphics[width=4.5in]{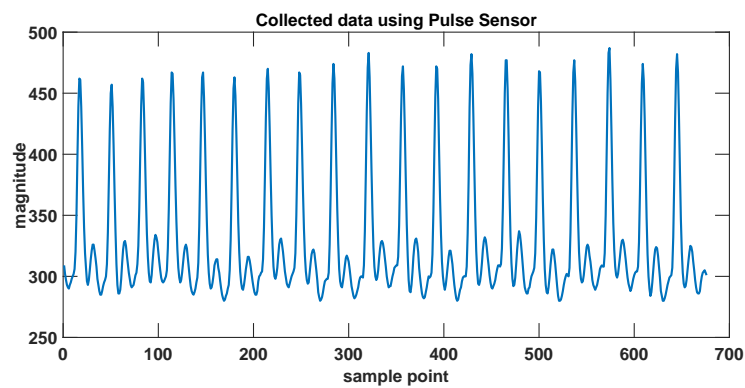}}
  {\caption{Collected ECG using Pulse Sensor.}\label{fig:PulseSensorECG}}
\end{figure}

In the second cross-database simulation, real-time ECG data is collected using an off-the-shelf plug-and-play heart-rate sensor, which is based on Arduino Uno \cite{Ahamed_2015}. This module is very popular among students, artists, athletes, and game \& mobile developers who want to incorporate live heart-rate data into their projects. The front- and back-side of this sensor are shown in Fig. \ref{fig:FrontBackSensor}. In the module, a sensor clip is tight to a fingertip or earlobe, while the other end is connected to the Arduino. The sensor clip collects the ECG data \cite{Ufoaroh_2015}.  It also includes an open-source monitoring app that graphs your pulse in real-time. The pulse sensor kit includes a 24-inch color-coded cable with header connectors, an ear clip, two velcro-dots, and three transparent stickers as shown in Fig. \ref{fig:FrontBackSensor2}. The collected ECG signal from the pulse sensor looks like as shown in Fig. \ref{fig:PulseSensorECG}. Using two pulse sensor modules, we can get two channels of the self-collected signal to mimic V2, and V5 leads. Then to make our model work, the signals are re-sampled at 360 samples per second. We have collected ECG signals from four different people for ten seconds. They do not have heart-related conditions. Using the SVM learning model trained by the SPHC arrhythmia database, all heartbeats extracted from these four normal people were classified as normal beats. 


\section{Conclusion}\label{Sec:Con}
Using SPHC database, our proposed ML model can predict different kinds of ECG heartbeats from different databases. The MIT-BIH database is a typical database for years of researches in arrhythmia classification \cite{Kin_2009}. Although this database has huge amount of data and many kinds of arrhythmia, it contains only ECG signals from 48 patients. So we choose this database to test our built ML model. The result shows that for normal heartbeats, predict recall reaches 100\% and for LBBB beats, the predict recall reaches 98.4\%. It shows a pretty high accuracy in classification of these two kinds of ECG heartbeats. Also we are expecting more cross-database simulations to generalize our ECG classification systems. 
Moreover, our realized system will be tested on heart patient in the future work.

\section*{acknowledgement}
The authors would like to thank the KAUST Smart Health Initiative for supporting this work.

\end{document}